\documentstyle[amssymb,aps,epsf,twocolumn]{revtex}

\def\be{\begin{equation}}
\def\ee{\end{equation}}
\def\bea{\begin{eqnarray}}
\def\eea{\end{eqnarray}}
\def\nn{\nonumber\\}

\begin{document}

\twocolumn[\hsize\textwidth\columnwidth\hsize\csname
@twocolumnfalse\endcsname

\title{Magnetization jump in the $XXZ$ chain with next-nearest-neighbor exchange}
\author{A. A. Aligia}
\address{Comisi\'{o}n Nacional de Energ{\'{\i }}a At\'{o}mica,\\
Centro At\'{o}mico Bariloche and Instituto Balseiro, 8400 S.C. de Bariloche,%
\\
Argentina}
\maketitle

\begin{abstract}
We study the dependence of the magnetization $M$ with magnetic field $B$ at
zero temperature in the spin-1/2 $XXZ$ chain with nearest-neighbor (NN) $%
J_{1}$ and next-NN $J_{2}$ exchange interactions, with anisotropies $\Delta
_{1}$ and $\Delta _{2}$ respectively. The region of parameters for which a
jump in $M(B)$ exists is studied using numerical diagonalization, and
analytical results for two magnons on a ferromagnetic background in the
thermodynamic limit. We find a line in the parameter space ($%
J_{2}/J_{1},\Delta _{1}/J_{1},\Delta _{2}/J_{1}$) (determined by two simple
equations) at which the ground state is highly degenerate. $M(B)$ has a jump
near this line, and at or near the isotropic case with ferromagnetic $J_{1}$
and antiferromagnetic $J_{2}$, with $|J_{2}/J_{1}|\sim 1/4$. These results
are relevant for some systems containing CuO chains with edge-sharing CuO$%
_{4}$ units.
\end{abstract}


\vskip2pc]
\narrowtext

\section{Introduction}

In recent years quantum spin chains and ladders have been a subject of great
experimental and theoretical interest. Quasi one-dimensional magnetic
systems have been identified and studied experimentally \cite
{mot,mat,col,john}, and new theoretical studies of the spin-1/2 $XXZ$ chain
with nearest-neighbor (NN) and next-NN exchange coupling (which is
equivalent to a two-leg zig-zag ladder) were presented \cite
{john,bur,whi,NGE,chp,ger,hir1,hir2,inco,itoi,hiki}. The Hamiltonian is:

\begin{eqnarray}
H &=&\sum_{i}[J_{1}(S_{i}^{x}S_{i+1}^{x}+S_{i}^{y}S_{i+1}^{y}+\Delta
_{1}S_{i}^{z}S_{i+1}^{z})  \nonumber \\
&&+J_{2}(S_{i}^{x}S_{i+2}^{x}+S_{i}^{y}S_{i+2}^{y}+\Delta
_{2}S_{i}^{z}S_{i+2}^{z})].  \label{h1}
\end{eqnarray}
One of the properties of this model is that the magnetization as a function
of applied magnetic field $M(B)$ displays a jump for certain parameters \cite
{ger,hir2}. This phenomenon called metamagnetic transition was observed for
example in TmSe \cite{bje}, Fe$_{x}$Mn$_{1-x}$TiO$_{3}$ \cite{ito}, Tb$%
_{1-x} $Sc$_{x}$Mn$_{2}$ \cite{lel}, and the quasi one-dimensional compound
Ba$_{3}$Cu$_{2}$O$_{4}$Cl$_{2}$ \cite{eck}.

Previous studies of metamagnetism in the model were restricted to $\Delta
_{1}=\Delta _{2}=\Delta $ \cite{ger,hir2}. The results show that a jump in $%
M(B)$ is not possible if $\Delta >\Delta _{c}$. Following the methods
explained in section III, we have determined $\Delta _{c}=(-5+\sqrt{17}%
)/4\cong -0.22$ \cite{icm}. In principle, the requirement of an opposite
sign for the $z$ component of the exchange than for the other two seems
unrealistic. However, for the Hamiltonian Eq. (\ref{h1}), a negative $\Delta
_{1}$ with positive $J_{1}$ is equivalent to a positive $\Delta _{1}$ with
negative $J_{1}$, since a rotation of every second spin in $\pi $ around the 
$z$ axis changes the sign of the $x$ and $y$ components of $J_{1}$. Thus, $%
\Delta _{1}<\Delta _{c}$ does not necessarily mean a large anisotropy of $%
\Delta _{1}$. Instead, a negative $\Delta _{2}<\Delta _{c}$ is a very large
anisotropy of $J_{2}$ and seems difficult to find in real systems.

In this work we extend the previous searches of metamagnetic transitions to
arbitrary values of $\Delta _{1}$ and $\Delta _{2}$, concentrating our study
in $|\Delta _{1}|\leq 1$ and $0\leq \Delta _{2}\leq 1$, and particularly
near the isotropic case of ferromagnetic $J_{1}$ and antiferromagnetic $%
J_{2} $. Several systems containing CuO chains with edge-sharing CuO$_{4}$
units, like La$_{6}$Ca$_{8}$Cu$_{24}$O$_{41}$, Li$_{2}$CuO$_{2}$, and Ca$%
_{2} $Y$_{2} $Cu$_{5}$O$_{10}$, are expected to lie near this limit \cite
{miz}. In these systems, the Cu-O-Cu angle $\theta $ of the CuO chains is
near $90^{\circ }$ and therefore, the usual antiferromagnetic NN exchange is
largely frustrated. As a consequence of virtual processes in which the Hund
rules exchange integral at the O atoms play a role, $J_{1}$ becomes small
and ferromagnetic. We estimate the critical field $B_{c}$ at which a jump or
an abrupt increase in $M(B)$ should exist in these compounds.

The outline of the paper is as follows. In section II we explain how we
determine the regions of parameters for which a metamagnetic transition is
expected. Section III contains analytical results for the onset of bound
states of two magnons in a ferromagnetic background. These results and
numerical ones in chains of 20 sites are used in section IV to present phase
diagrams in which the boundaries of the metamagnetic regions are shown. This
section includes numerical results for the ground state energy per site as a
function of magnetic field $E(M)$, curves of $M(B)$, and critical field $%
B_{c}$ as a function of $\alpha =J_{2}/J_{1}$ for $\Delta _{2}=-\Delta _{1}=1
$. Section V contains a more detailed numerical study of this isotropic case
for $\alpha $ slightly larger than 1/4. Section VI is a summary and
discussion.

\section{Conditions for the existence of a jump in $M(B)$}

For the sake of clarity, we anticipate some of our numerical results for the
dependence of the ground state energy per site $E$ as a function of
magnetization $M=S^{z}/L$, where $S^{z}=\sum_{i}S_{i}^{z}$ is the $z$
component of the total spin and $L$ is the number of sites. They are shown
in Fig. 1. A large portion of the curve $E(M)$ has negative curvature. These
points are actually not accessible thermodynamically. The dashed straight
line is tangent to $E(M)$ at the two points $M_{1}=0.108$ and $M_{2}=0.5$
(the Maxwell construction). For all $M$ in the interval $(M_{1},M_{2})$, it
is energetically more favorable for the system to phase separate into a
fraction $x=(M-M_{1})/(M_{2}-M_{1})$ with magnetization $M_{2}$ and a
fraction $1-x$ with magnetization $M_{1}$. The energy of the mixture is
represented by the dashed line. If a magnetic field $B$ is applied to the
system, the equilibrium magnetization is determined minimizing the Helmholz
transform $G=E-MB$, what leads to the condition $B=\partial E/\partial M$.
From inspection of Fig. 1, we see that $M(B)$ increases with $B$, until it
reaches the critical value $B_{c}=[E(M_{2})-E(M_{1})]/(M_{2}-M_{1})$ ($%
=0.00277J_{1}$ for the case of Fig. 1). At $B=B_{c}$, $M(B)$ suddenly jumps
from $M_{1}$ to $M_{2}$. If $M_{2}$ is smaller than the saturation
magnetization, $M$ increases further for $B>B_{c}$, but we have not found
this situation in the present model.

\begin{figure}
\narrowtext
\epsfxsize=3.5truein
\vbox{\hskip 0.05truein \epsffile{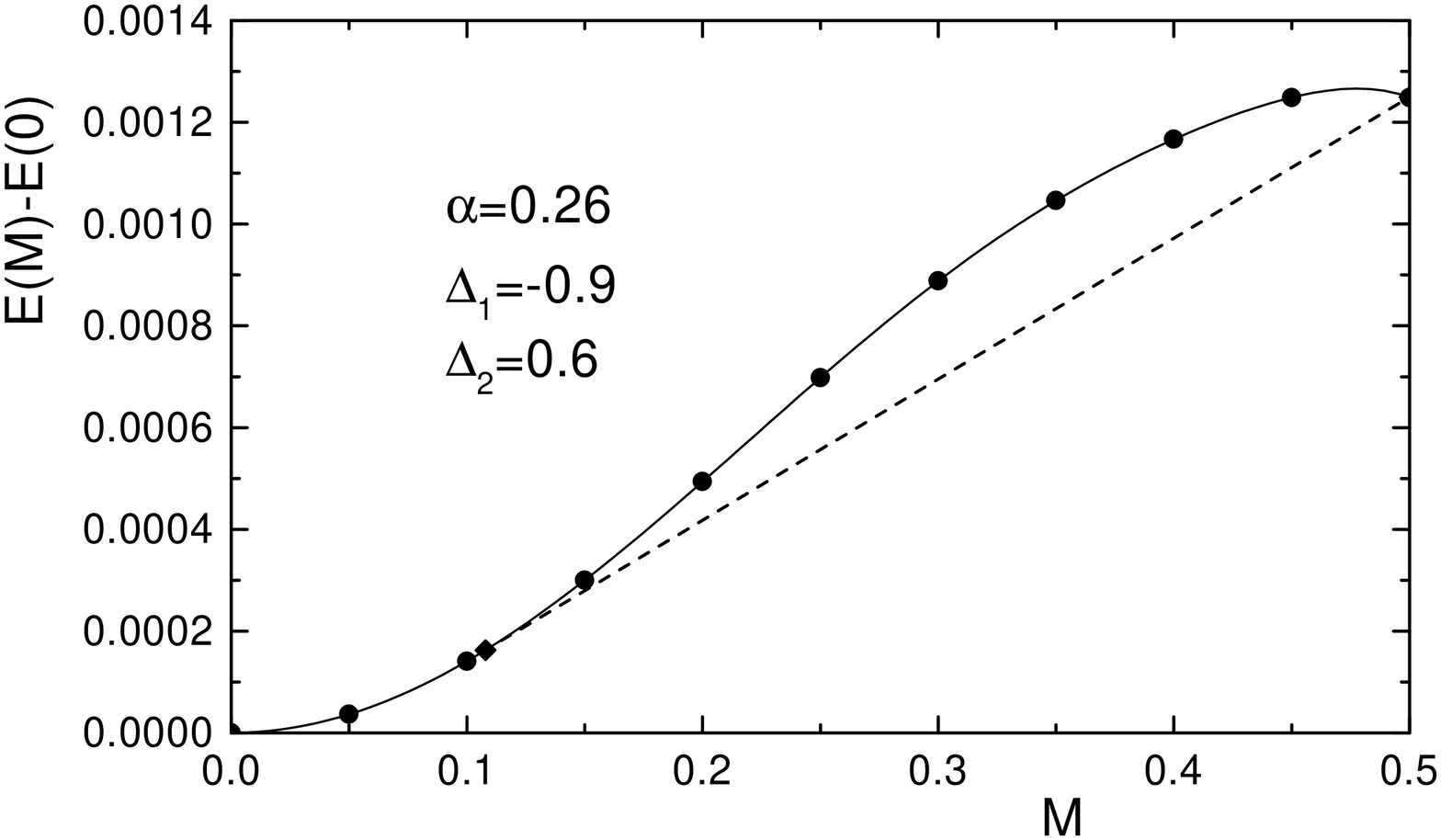}}
\medskip
\caption{Energy per site as a function of total $z$ component of spin
per site for a chain of 20 sites (solid circles). The full line is a
polynomial fit. Dashed line and diamonds correspond to the Maxwell
construction. Parameters are $J_{1}=1$, $\alpha =J_{2}/J_{1}=0.25$, $\Delta
_{1}=-0.9$ and $\Delta _{2}=0.6$.}
\end{figure}

From the above description, one can see that for a metamagnetic transition
to occur at very low temperatures, $E(M)$ should satisfy two conditions: 1) $%
\partial ^{2}E/\partial M^{2}<0$ in a finite interval of values of $M$. 2) $%
E(M_{2})>E(M_{1})$, where $M_{1}<M_{2}$ are determined by the Maxwell
construction. From the general behavior of $E(M)$ for the case $\Delta
_{1}=\Delta _{2}=\Delta $ , Gerhardt {\it et al. }\cite{ger} have found that
when metamagnetism exists, $M_{2}=1/2$ and the condition 2 ceases to be
satisfied when $M_{1}=0$. More precisely, from their finite-size results for 
$E(M,\alpha ,\Delta )$, with $\alpha =J_{2}/J_{1}$, they obtained a critical
value of $\Delta $ ($\Delta ^{f}(\alpha )$) from the equation $E(0,\alpha
,\Delta ^{f})=E(1/2,\alpha ,\Delta ^{f})$. For $\Delta <\Delta ^{f}$ the
system is a fully spin polarized ferromagnet even at $B=0$. Fortunately, the
results for $\Delta ^{f}$ do not show a significant size dependence and it
is accurately determined in chains of 18 sites. Another critical value $%
\Delta ^{a}(\alpha )$ was obtained from the condition $\partial
^{2}E/\partial M^{2}|_{M=1/2}=0$. For $\Delta >\Delta ^{a}$ the curvature $%
\partial ^{2}E/\partial M^{2}$ is positive for all $M$. The curvature at $%
M_{2}$ was calculated numerically using:

\begin{eqnarray}
\partial ^{2}E/\partial M^{2}|_{M=1/2} &=&\lim_{L\rightarrow \infty
}L^{2}[E(1/2)+E(1/2-2/L)  \nonumber \\
&&-2E(1/2-1/L)],  \label{c1}
\end{eqnarray}
in a periodic chain. In contrast to the case of $\Delta ^{f}$, and even
taking $L=50,$ the result has some finite-size effects \cite{ger}. This is
in part due to the fact (explained in the next section) that the ground
state near $M=1/2$ becomes incommensurate for sufficiently large $\alpha $,
and these wave vectors cannot be represented in small chains if periodic
boundary conditions are used \cite{inco}. From the numerical solution of the
problem of two spin excitations on the ferromagnetic state for $L\rightarrow
\infty $, more accurate values of $\Delta ^{a}(\alpha )$ were obtained
recently for $\alpha \leq 1/2$ \cite{hir2}. In the region of the $(\alpha
,\Delta )$ plane where $\Delta ^{f}(\alpha )<\Delta <\Delta ^{a}(\alpha )$ a
metamagnetic transition occurs in the model \cite{ger,hir2}.

Relaxing the condition $\Delta _{1}=\Delta _{2}=\Delta $, and keeping for
example $\Delta _{2}$ fixed, limiting values of $\Delta _{1}$ ($\Delta
_{1}^{f}(\alpha ,\Delta _{2})$ and $\Delta _{1}^{a}(\alpha ,\Delta _{2})$)
can be defined in the same way and we expect in principle that magnetization
jumps will be found if $\Delta _{1}^{f}(\alpha ,\Delta _{2})<\Delta
_{1}<\Delta _{1}^{a}(\alpha ,\Delta _{2})$. The same is true interchanging $%
\Delta _{1}$ and $\Delta _{2}$. Fortunately, as described in the next
section, the $\Delta _{i}^{a}$ are given by analytical expressions, which
are very simple near $\alpha =1/4$. The $\Delta _{i}^{f}$ are determined by
numerical diagonalization of chains with $L=20$ and the results are given in
section IV.

\section{The two-magnon problem}

In this section, we derive analytical expressions for the upper boundaries
of the expected metamagnetic region ($\Delta _{1}^{a}(\alpha ,\Delta _{2})$
or $\Delta _{2}^{a}(\alpha ,\Delta _{1})$), looking for bound states of two
spin flips on a ferromagnetic background in an infinite chain. This is a
two-body problem which due to translational invariance can be mapped into a
single-body one. Similar problems were solved for example to determine
exactly metal-insulator boundaries in generalized Hubbard models \cite{ov}.
For $\Delta _{1}=\Delta _{2}=\Delta $, this problem has been solved
numerically for $L\sim 150$ by Cabra {\it et al.} \cite{chp} and for $%
L\rightarrow \infty $ and $\alpha \leq 1/2$ by Hirata \cite{hir2}. Simple
analytical expressions are given in Ref. \cite{icm}.

We use a Jordan-Wigner transformation $S_{j}^{+}=c_{j}^{\dagger }\exp (i\pi
\sum_{l<j}n_{l})$, $S_{j}^{-}=(S_{j}^{+})^{\dagger }$, $S_{j}^{z}=n_{j}-1$,
with $n_{j}=c_{j}^{\dagger }c_{j}$, to express the spin operators in terms
of spinless fermions. Calling $\alpha =J_{2}/J_{1}$, setting $J_{1}=1$ as
the unit of energy, and substracting the constant $LE(1/2)$ (with $%
E(1/2)=(\Delta _{1}+\alpha \Delta _{2})/4$), the Hamiltonian Eq. (\ref{h1})
takes the form:

\begin{eqnarray}
H &=&\sum_{i}[(-\Delta _{1}-\alpha \Delta _{2})n_{i}+\frac{1}{2}%
(c_{i+1}^{\dagger }c_{i}+\alpha c_{i+2}^{\dagger }c_{i}+\text{H.c.}) 
\nonumber \\
&&+\Delta _{1}n_{i}n_{i+1}+\alpha \Delta _{2}n_{i}n_{i+2}-\alpha
(c_{i+2}^{\dagger }n_{i+1}c_{i}+\text{H.c.})].  \label{h2}
\end{eqnarray}
After Fourier transform $c_{j}=\sqrt{1/L}\sum_{q}e^{iqj}c_{q}$, the model
can be written as:

\begin{eqnarray}
H &=&\sum_{q}\epsilon _{q}c_{q}^{\dagger }c_{q}+\frac{4}{L}%
\sum_{K}\sum_{q,q^{\prime }>0}[(\Delta _{1}+2\alpha \cos K)\sin q\sin
q^{\prime }  \nonumber \\
&&+\alpha \Delta _{2}\sin 2q\sin 2q^{\prime }]c_{\frac{K}{2}-q^{\prime
}}^{\dagger }c_{\frac{K}{2}+q^{\prime }}^{\dagger }c_{\frac{K}{2}+q}c_{\frac{%
K}{2}-q},  \label{h3}
\end{eqnarray}
with $\epsilon _{q}=-\Delta _{1}-\alpha \Delta _{2}+\cos q+\alpha \cos 2q$.

The two-magnon eigenstates of total momentum $K$ can be written in the form $%
|\psi (K)\rangle =\sum_{q>0}A_{q}c_{\frac{K}{2}-q}^{\dagger }c_{\frac{K}{2}%
+q}^{\dagger }|0\rangle $. Replacing this into the Schr\"{o}dinger equation $%
H|\psi \rangle =\lambda |\psi \rangle $, one obtains:

\begin{equation}
A_{q}=\frac{-4[(\Delta _{1}+2\alpha \cos K)S_{1}(K)\sin q+\alpha \Delta
_{2}S_{2}(K)\sin 2q]}{\epsilon _{\frac{K}{2}-q}+\epsilon _{\frac{K}{2}%
+q}-\lambda },  \label{a}
\end{equation}
where

\begin{equation}
S_{n}(K)=\frac{1}{L}\sum_{q>0}A_{q}\sin nq.  \label{s}
\end{equation}
Replacing Eq. (\ref{a}) into Eq. (\ref{s}) leads to two homogeneous
equations for the two unknown sums $S_{1}$ and $S_{2}$. For fixed values of
the parameters and $K$, the condition of vanishing secular determinant
determines the allowed eigenvalues $\lambda $. Since we are looking for the
condition on the parameters for the onset of a bound state (when the right
hand side of Eq. (\ref{c1}) becomes negative), we set $\lambda $ slightly
below two times the minimum one-magnon energy: $\lambda =2\min (\epsilon
_{q})-\eta $, where $\eta $ is a positive infinitesimal energy. With this
value of $\lambda $, the condition of vanishing secular determinant in the
thermodynamic limit $L\rightarrow \infty $ takes the form:

\bea
4(\Delta _{1}+2\alpha \cos K)[I_{0}+16\alpha \Delta
_{2}(I_{0}I_{2}-I_{1}^{2})] \nn
+16\alpha \Delta _{2}I_{2}+1=0,  \label{det}
\eea
where the integrals $I_{n}(\alpha ,K)$ ($n=0,1,2$) are:

\begin{equation}
I_{n}=\frac{1}{2\pi }\int_{0}^{\pi }\frac{dq\sin ^{2}q\cos ^{n}q}{4\alpha
\cos K\cos ^{2}q+2\cos \frac{K}{2}\cos q+c(\alpha ,K)+\eta },  \label{integ}
\end{equation}
with $\eta \rightarrow 0$ and

\begin{eqnarray}
c(\alpha ,K) &=&2(1-\alpha -\alpha \cos K)\text{, \qquad }\alpha \leq 1/4 
\nonumber \\
c(\alpha ,K) &=&2\alpha +\frac{1}{4\alpha }-2\alpha \cos K\text{, \thinspace 
}\alpha \geq 1/4\text{.}  \label{ci}
\end{eqnarray}
The change of the expression for $c(\alpha ,K)$ at $\alpha =1/4$ is due to
the change in the wave vector which minimizes $\epsilon _{q}$ from $q=\pi $
for $\alpha \leq 1/4$ to $q=\pm \arccos (-1/(4\alpha ))$ for $\alpha \geq 1/4
$. The integrals can be solved decomposing the integrand into a sum of
expressions with denominators linear in $\cos q$ and using \cite{gr}

\begin{equation}
\frac{1}{2\pi }\int_{0}^{\pi }\frac{dq\sin ^{2}q}{a+b\cos q}=\frac{a}{2b^{2}}%
\left( 1-\sqrt{1-\frac{b^{2}}{a^{2}}}\right) .  \label{gr}
\end{equation}

For each value of $\alpha $, $K$ and $\Delta _{1}$, Eq. (\ref{det}) is a
linear equation in $\Delta _{2}$. The searched upper boundary $\Delta
_{2}^{a}(\alpha ,\Delta _{1})$ is determined choosing the value of $K$ which
leads to the highest root of Eq. (\ref{det}). For $\Delta _{2}<\Delta
_{2}^{a}(\alpha ,\Delta _{1})$, the curvature Eq. (\ref{c1}) is negative.
The same is true interchanging $\Delta _{1}$ and $\Delta _{2}$. If $\Delta
_{1}=\Delta _{2}=\Delta $ is taken \cite{ger,hir2}, the highest root of the
quadratic Eq. (\ref{det}) determines $\Delta ^{a}(\alpha )$, since there is
at least one bound state for $\Delta <\Delta ^{a}(\alpha )$.

For $\alpha \leq 1/4$, the total wave vector $K$ which first leads to a
bound state is $K=0$. In the sector of two particles, this is also the wave
vector of the ground state of the non-interacting part of the Hamiltonian in
the fermionic representation (Eq. (\ref{h2}) or (\ref{h3})). Using Eqs. (\ref
{integ}), (\ref{ci}) and (\ref{gr}), we obtain after some algebra:

\begin{eqnarray}
I_{0} =\frac{r-1}{8\alpha }\text{, }I_{1}=\frac{1-(1-2\alpha )r}{16\alpha
^{2}}\text{, } \nn 
I_{2}=\frac{\alpha -1}{16\alpha ^{2}}+\left( \frac{1}{2\alpha }
-1\right) ^{2}I_{0}\text{,}  \nn
\text{ with }r =(1-4\alpha )^{-1/2}\text{, }(\alpha <1/4).  \label{i1}
\end{eqnarray}
Eqs. (\ref{det}) and (\ref{i1}) define $\Delta _{2}^{a}(\alpha ,\Delta _{1})$
and $\Delta _{1}^{a}(\alpha ,\Delta _{2})$ for $\alpha <1/4.$

For $\alpha >1/4$, the ground state of the non-interacting fermionic
Hamiltonian for two particles is degenerate with total wave vector $K=0$ or $%
K=\pm K_{i}$ with $K_{i}=2\arccos (-1/(4\alpha ))$. For $\Delta _{1}=\Delta
_{2}$, we have found that the wave vector of the ground state of the
two-magnon problem for parameters near the onset of a bound state is $K=0$
for $\alpha \leq 1/2$. At $\alpha =1/2$ it jumps to $\pm K_{i}$. From
numerical investigations in finite systems (with $L\sim 20$), using twisted
boundary conditions to allow all possible wave vectors \cite{inco}, we find
that for values of $\alpha >\alpha _{w}\sim 1$, the ground state wave vector
deviates continuously from $\pm K_{i}$. Comparison of the numerical results
for $\Delta ^{a}(\alpha )$ (with $L\sim 50$) of Ref. \cite{ger} with our
analytical ones assuming $K=\pm K_{i}$, gives $\alpha _{w}\cong 0.77$. In
the region $\alpha >\alpha _{w}$, it seems not possible to find analytical
results for $\Delta ^{a}(\alpha )$. However, taking $K=K_{i}$ leads to a
reasonable lower bound, since $K$ is not too different from $K_{i}$ (in any
case, as we shall see in section V, Eq. (\ref{c1}) ceases to be valid for $%
\alpha \geq \alpha _{w}$). In the general case $\Delta _{1}\neq \Delta _{2}$%
, the maximum between the results for $\Delta _{n}$ assuming $K=0$ or $%
K=K_{i}$, also gives a lower bound for $\Delta _{2}^{a}(\alpha ,\Delta _{1})$
and $\Delta _{1}^{a}(\alpha ,\Delta _{2})$. We restrict the calculation to
these values of $K$. For $K=0$ and $\alpha \geq 1/4$, the integrals Eq. (\ref
{integ}) diverge for $\eta \rightarrow 0$, and Eq. (\ref{gr}) has complex
coefficients for finite $\eta $. The equation for the $\Delta _{n}$ is
obtained after a careful limiting procedure of Eq. (\ref{det}) with adequate
choice of the branch of the root in Eq. (\ref{gr}). Physically this means
simply to consider values of $\Delta _{n}$ slightly smaller than the
critical ones, in such a way that they lead to a finite binding energy $\eta 
$, and then take the limit $\eta \rightarrow 0$. The final results turns out
to be very simple:

\begin{eqnarray}
\frac{\Delta _{1}\Delta _{2}+\Delta _{1}}{2\alpha }+\Delta _{2}(1+\frac{1}{%
8\alpha ^{2}})+1 &=&0\text{,}  \nonumber \\
(\alpha &\geq &1/4\text{, }K=0).  \label{det2}
\end{eqnarray}
As an example, this equation is satisfied for $\alpha =1/4$, $\Delta
_{1}=-0.9$, $\Delta _{2}=2/3$. Lowering $\Delta _{2}$ a little bit, $E(M)$
displays the behavior required for a jump in $M(B)$ (see Fig. 1).

Finally, for $\alpha \geq 1/4$ and $K=K_{i}$, the integrals take the form:

\begin{eqnarray}
I_{0} &=&\frac{1}{8\alpha B}\left( 1-\frac{1}{4\alpha \sqrt{A}}\right) \text{%
, }I_{1}=\frac{1}{32(\alpha B)^{2}}\left( \frac{1}{2\alpha }-\frac{1}{\sqrt{A%
}}\right) \text{,}  \nonumber \\
I_{2} &=&\frac{1}{16\alpha }\left( \frac{1}{A}-\frac{1}{B}+\frac{2\alpha -%
\sqrt{A}}{2\alpha AB^{3}}\right) \text{, with }  \nonumber \\
A &=&1-\frac{1}{16\alpha ^{2}}\text{, }B=1-\frac{1}{8\alpha ^{2}}\text{, }%
(\alpha \geq 1/4\text{, }K=K_{i}).  \label{i3}
\end{eqnarray}
It remains to establish when $\Delta _{i}^{a}$ are determined by Eq. (\ref
{det2}), or Eq. (\ref{i3}) and Eq. (\ref{det}) with $\cos K=1/(8\alpha
^{2})-1$. Taking for instance $\Delta _{1}$ fixed, equating the values of $%
\Delta _{2}$ obtained from both expressions, we obtain:

\begin{equation}
\alpha =-\frac{1}{4\Delta _{1}}\text{, }\Delta _{2}=-1+2\Delta _{1}^{2}\text{%
, }(-1\leq \Delta _{1}\leq 0).  \label{line}
\end{equation}
For smaller values of $\alpha $, $\Delta _{2}^{a}(\alpha ,\Delta _{1})$ is
determined by Eq. (\ref{det2}). Similarly, for a given value of $\Delta _{2}$%
, Eq. (\ref{det2}) gives $\Delta _{1}^{a}(\alpha ,\Delta _{2})$ if $\alpha
\leq [(1+\Delta _{2})/2]^{-1/2}/4$, while for larger values of $\alpha $,
Eqs. (\ref{det}) and (\ref{i3}) should be used.

Eq. (\ref{line}) defines a particular line in the parameter space $(\alpha
,\Delta _{1},\Delta _{2})$, which as will be seen in the next section, plays
an important role in our study. From the analytical results of this section,
we see that at this line, not only there is a degeneracy in the lowest lying
eigenstates in the two-magnon sector (for $M=$ $1/2-2/L$ with $L\rightarrow
\infty $) between the wave vectors $K=0$ and $K=\pm K_{i}$, but also $%
\lim_{L\rightarrow \infty }L(E(1/2-1/L)-E(1/2))=\epsilon _{\arccos (\Delta
_{1})}=0$ and $\lim_{L\rightarrow \infty }L(E(1/2-2/L)-E(1/2))=0$. In other
words $E(M)$ is independent of $M$ for $M\sim 1/2$ in the thermodynamic
limit.

\section{Phase diagram for metamagnetic transitions}

In this section, we delimit the region of parameters inside which a jump in
the function $M(B)$ is expected. Specifically for each $\alpha $ and $\Delta
_{1}$, we determine by numerical diagonalization of chains of 20 sites the
lower boundary $\Delta _{2}^{f}$ of the metamagnetic region (from $%
E(1/2)=E(0)$), while the upper boundary $\Delta _{2}^{a}$ (or a lower bound
of it for large $\alpha $) is taken from the expressions of the previous
section. The same is done interchanging $\Delta _{1}$ and $\Delta _{2}$.
This study is complemented by numerical calculations of the ground-state
energy as a function of magnetization $E(M)$ (from which $M(B)$ can be
derived), and the critical field $B_{c}$ for several parameters inside the
region of interest.

\begin{figure}
\narrowtext
\epsfxsize=3.5truein
\vbox{\hskip 0.05truein \epsffile{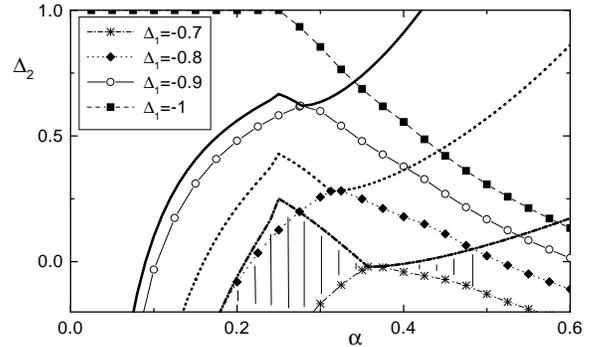}}
\medskip
\caption{Boundaries of the region of expected metamagnetic transitions
in the ($\alpha ,\Delta _{2}$) plane, for different values of $\Delta _{1}$.
This region is shown dashed for $\Delta _{1}=-0.7$. The symbols (joined by
thin lines) indicate the lower boundary $\Delta _{2}^{f}$ calculated
numerically in chains of 20 sites. The upper boundary $\Delta _{2}^{a}$
(thick lines) is taken from the analytical results of section III.}
\end{figure}

In Fig. 2 we show $\Delta _{2}^{f}$ (discrete points) and $\Delta _{2}^{a}$
(thick curve above it) as a function of $\alpha $ for several values of $%
\Delta _{1}$. The region of interest $\Delta _{2}^{f}<$ $\Delta _{2}<\Delta
_{2}^{a}$ is shown by vertical dashed lines for $\Delta _{1}=-0.7$. Clearly
the upper boundary $\Delta _{2}^{a}$ displays two kinks as a result of the
change in the ground state wave vector for one magnon at $\alpha =1/4$, or
two magnons at $\alpha =-1/(4\Delta _{1})$, as described in the previous
section. $\Delta _{2}^{a}$ is given by Eqs. (\ref{i1}) and (\ref{det}) with $%
K=0$ for $\alpha <1/4$, by Eq. (\ref{det2}) between both kinks, and (as a
lower bound) by Eqs. (\ref{i3}) and (\ref{det}) with $\cos K=1/(8\alpha
^{2})-1$ for $\alpha >-1/(4\Delta _{1})$ respectively. It is remarkable that
at the second kink (determined by Eqs. (\ref{line})), where the wave vector
of the two-magnon ground state jumps from $K=0$ to $K=\pm K_{i}$ and $%
\partial E/\partial M|_{M=1/2}=\partial ^{2}E/\partial M^{2}|_{M=1/2}=0$ for 
$L\rightarrow \infty $ according to the results of the previous section,
also $E(1/2)=E(0)$ (at least within the accuracy of the Lanczos
diagonalization an independently of system size). Actually $E(M)$ is quite
flat at this point. Specifically, we find by numerical diagonalization that 
{\em if and only if }the energy for each value of $S^{z}$ is minimized with
respect to the optimum twisted boundary conditions \cite{inco}, $E(M)$
becomes independent of $M$ at this point, within our precision of $10^{-9}$.
All wave vectors at the minimum become incommensurate except for $S^{z}=0$
and $S^{z}=L/2$. For $S^{z}=L/2-2$, we find that the minimum is at wave
vector $K=K_{i}$, while the energy at $K=0$ is of the order of $10^{-5}$
above the minimum energy for $L=20$, due to finite size effects. 

Particular cases of degeneracy at the line determined by Eqs. (\ref{line})
are already known. The degeneracy at the point $\alpha =1/4$, $\Delta _{1}=-1
$, $\Delta _{2}=1$ was studied by Hamada {\it et al.}, who also have shown
that the ground state in the sector of total spin $S=0$ is a
resonance-valence-bond state involving singlet pairs at all distances \cite
{ham}. The degeneracy at the Majumdar-Ghosh point $\alpha =1/2$, $\Delta
_{1}=\Delta _{2}=-1/2$, was discussed by Gerhardt {\it et al.} \cite{ger}.
For $\Delta _{1}\rightarrow 0$, this point moves towards two decoupled
ferromagnetic Heisenberg chains (see Eqs. (\ref{line})), for which $E(M)$ is
constant.

This special point where $\Delta _{2}^{f}$ and $\Delta _{2}^{a}$ coincide
separates the region of expected metamagnetism in two zones. The one at the
left of the point shrinks and moves towards $\Delta _{2}=1$ as $\Delta _{1}$
decreases approaching the isotropic limit $\Delta _{1}=-1$, where it
disappears. The right zone also displaces towards $\Delta _{2}=1$, but
increases as $\Delta _{1}$ moves to $-1$, suggesting that a jump in $M(B)$
is also possible in the isotropic limit $\Delta _{2}=-\Delta _{1}=1$, if $%
\alpha >1/4$. Fig. 3 shows the same two metamagnetic zones in the ($\alpha
,\Delta _{1}$) plane for several values of $\Delta _{2}$. The same
tendencies as before are observed as the isotropic limit $\Delta
_{2}=-\Delta _{1}=1$ is approached.

\begin{figure}
\narrowtext
\epsfxsize=3.5truein
\vbox{\hskip 0.05truein \epsffile{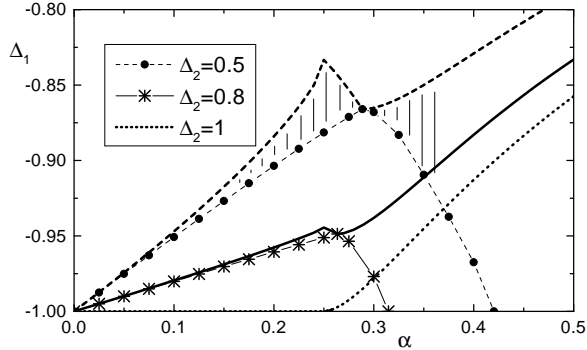}}
\medskip
\caption{Boundaries of the region of expected metamagnetic transitions
in the ($\alpha ,\Delta _{1}$) plane, for different values of $\Delta _{2}$.
This region is shown dashed for $\Delta _{2}=0.5$. The symbols (joined by
thin lines) indicate the lower boundary $\Delta _{1}^{f}$ calculated
numerically in chains of 20 sites. The upper boundary $\Delta _{1}^{a}$
(thick lines) is taken from the analytical results of section III.}
\end{figure}

In the zone of lower $\alpha $, the existence of a jump in $M(B)$ is
confirmed by numerical calculation of $E(M)$. An example was shown in Fig.
1. Instead, inside the right zone of expected metamagnetism (for $\alpha $
larger than the point of degeneracy at which $\Delta _{2}^{f}$ and $\Delta
_{2}^{a}$ meet), the situation is not so clear. As it is clear in Fig. 4 for 
$\alpha =1/2$, the   

\begin{figure}
\narrowtext
\epsfxsize=13. truecm
\vbox{\hskip 0.05truein \epsffile{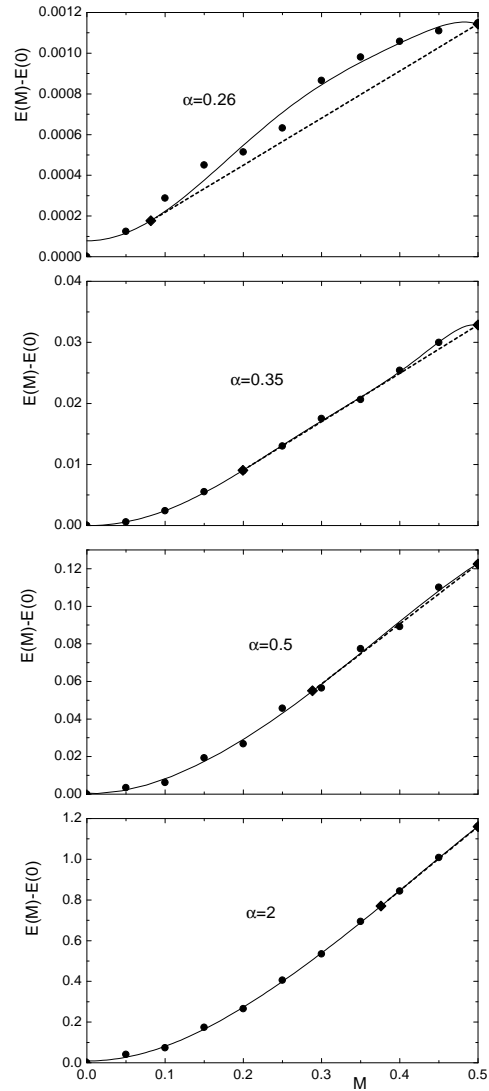}}
\medskip
\caption{Energy per site as a function of total spin per site for a
chain of 20 sites with $J_{1}=1$, $\Delta _{1}=-1$, $\Delta _{2}=1$, and
several values of $\alpha =J_{2}/J_{1}$. Full lines are polynomial fits (see
text). Dashed line and diamonds correspond to the Maxwell construction.}
\end{figure}

energy per site as a function of the $z$ component of the
total spin $S^{z}=\sum_{i}S_{i}^{z}=ML$, shows a significant even-odd effect
for small chains. This effect has been reduced minimizing $E(M)$ with
respect to the optimum twisted boundary conditions to allow for
incommensurate wave vectors. \cite{inco} Most of the resulting wave vectors
are incommensurate, particularly for odd $S^{z}$ and low $\alpha >1/4$. In
spite of this procedure, for $\alpha >0.4$, all $E(M)$ for odd $S^{z}$ seem
to be shifted to higher energies in comparison with the corresponding values
for even $S^{z}$. If this tendency persists in the thermodynamic limit
(keeping $L$ even) states with odd $S^{z}$ would not be accessible
thermodynamically, and the calculation of the previous section (based on Eq.
(\ref{c1})) would be irrelevant. In any case $E(M)$ is very flat for $%
0.3<M<0.5$ (see Fig. 4) and if the curvature at $M=1/2$ is positive, there
would be a steep increase in $M(B)$ for $M>0.3$, which is probably hard to
distinguish experimentally from a jump. The determination of the critical
value of $\alpha $ for which $\partial ^{2}E/\partial M^{2}|_{M=1/2}$
changes sign, is a difficult task which is postponed to the next section.

\begin{figure}
\narrowtext
\epsfxsize=3.5truein
\vbox{\hskip 0.05truein \epsffile{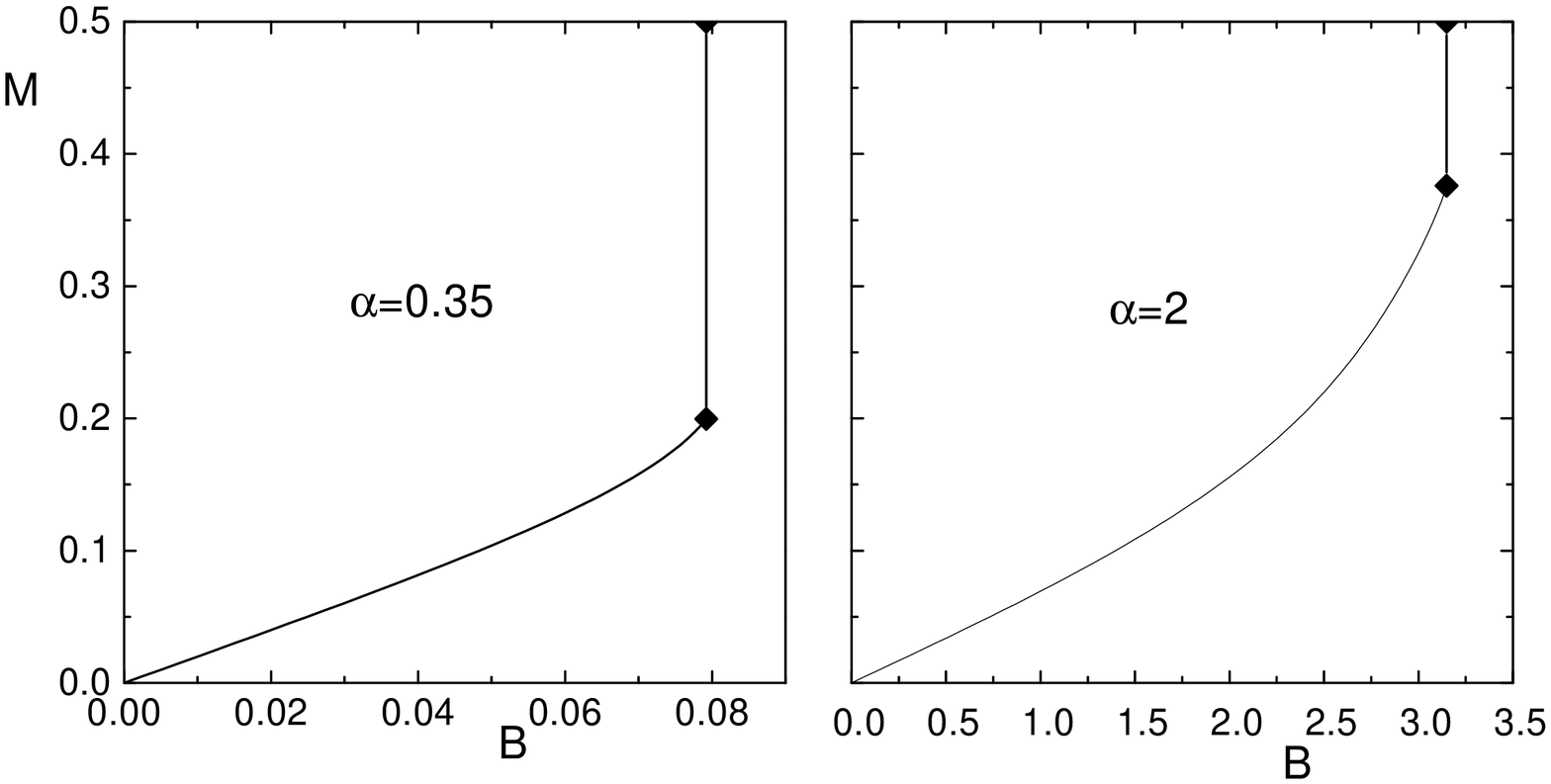}}
\medskip
\caption{Total spin per site as a function of magnetic field for $%
J_{1}=1$, $\Delta _{1}=-1$, $\Delta _{2}=1$, and two values of $\alpha $.}
\end{figure}

The continuous curves in Fig. 4 correspond to fits in the numerical data
using a polynomial of even powers of $M$ with about half as many parameters
as points to be fitted (to average the even-odd effect). These continuous
curves allowed us to perform the Maxwell construction analytically (dashed
lines and diamonds in Fig. 4), and to calculate $B(M)$ from $B=\partial
E/\partial M$ (see section II). The magnetization curve $B(M)$ is shown in
Fig. 5 for two values of $\alpha $, which are near those estimated for La$%
_{6}$Ca$_{8}$Cu$_{24}$O$_{41}$, and Ca$_{2}$Y$_{2}$Cu$_{5}$O$_{10}$
respectively. \cite{miz} While the details of the curve and the value of the
critical field at the jump $B_{c}$ (if it exists) depend on the fitting
procedure, the fact that there is an abrupt increase from $\sim 60\%$ to
100\% of maximum magnetization with a very small variation of magnetic field
is a genuine feature of the system. To estimate the variation of $B_{c}$
with $\alpha $ for the isotropic model $\Delta _{2}=-\Delta _{1}=1$, we have
calculated the average slope of $E(M)$ (minimized with respect to the
optimum twisted boundary conditions) between $M=0.3$ and $M=0.5$. The result
is shown in Fig. 6. Assuming $g=2$ for the gyromagnetic factor of the Cu
ions, and taking the values of $J_{1}$ and $\alpha $ estimated for La$_{6}$Ca%
$_{8}$Cu$_{24}$O$_{41}$, Li$_{2}$CuO$_{2}$, and Ca$_{2}$Y$_{2}$Cu$_{5}$O$%
_{10}$, \cite{miz} we obtain $B_{c}=$ 14, 40 and 65 Tesla respectively.

\begin{figure}
\narrowtext
\epsfxsize=3.5truein
\vbox{\hskip 0.05truein \epsffile{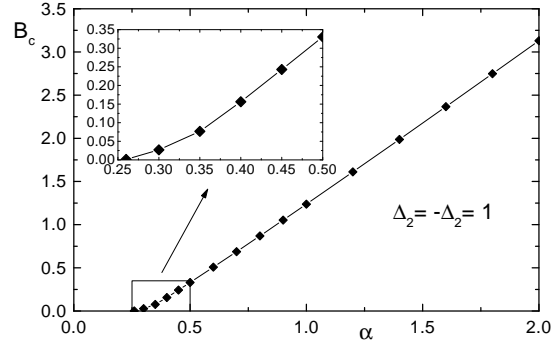}}
\medskip
\caption{Critical magnetic field as a function of $\alpha $ for $J_{1}=1
$, $\Delta _{1}=-1$, $\Delta _{2}=1$.}
\end{figure}

\section{Magnetization jump in the isotropic case}

The results of the previous section for $\Delta _{2}=-\Delta _{1}=1$, show
that there is an abrupt increase in $M(B)$ for $B\sim B_{c}(\alpha )$ near $%
M=1/2$, particularly for $1/4<\alpha \leq 1/2$. However, they are not enough
to establish the existence of a true jump. The calculation of $\partial
^{2}E/\partial M^{2}|_{M=1/2}$ becomes complicated by the fact that,
independently of system size (at least for $L\lesssim 40$ and near $M=1/2$),
only states with total spin $S=L/2-In_{\min }$ with $I$, $n_{\min }$
integers, but $n_{\min }>1$, can be of thermodynamic relevance. In other
words, the curve $E(M)$ near $M=1/2$ looks like a straight line of slope $%
B_{c}$ plus a periodic function with period $n_{\min }/L$ (see for example
Fig. 4 for $\alpha =0.5$, where $n_{\min }=2$; and Fig. 7 for $\alpha =0.3$,
where $n_{\min }=3$). If $\partial ^{2}E/\partial M^{2}|_{M=1/2}>0$, this
means (as stated in Ref. 8) that reducing the magnetic field from values
high enough to ensure saturation of the magnetization, the spins flip in
groups of $n_{\min }$. The physical reason of this behavior is not
completely clear. It seems that spin flips tend to bind in groups of $%
n_{\min }.$

As a consequence, Eq. (\ref{c1}) which could be calculated analytically
becomes invalid and should be replaced by::

\begin{eqnarray}
\partial ^{2}E/\partial M^{2}|_{M=1/2} &=&\lim_{L\rightarrow \infty }\frac{%
L^{2}}{n_{\min }^{2}}[E(1/2)+E(1/2-2n_{\min }/L)  \nonumber \\
&&-2E(1/2-n_{\min }/L)].  \label{c2}
\end{eqnarray}
While $n_{\min }=2$ for $\alpha >0.4$, $n_{\min }$ increases with decreasing 
$\alpha $, making larger system sizes necessary for an accurate evaluation
of Eq. (\ref{c2}). In addition, this equation cannot be used for values of $%
\alpha $ for which $n_{\min }$ is not well defined. For example $\alpha \sim
0.35$ (see Fig. 4) seems to correspond to a transition from $n_{\min }=2$ to 
$n_{\min }=3$ with lowering $\alpha $. Taking into account these
difficulties, we have chosen three values of $\alpha $ in the range $%
1/4<\alpha \leq 0.3$, with apparently well defined values of $n_{\min }$ (3,
4 and 5 for $\alpha =0.3$, 0.28 and 0.26 respectively, see Fig. 7), and have
calculated Eq. (\ref{c2}) as a function of system size. For $\alpha =0.3$,
for which the energy with six ($2n_{\min }$) spin flips should be
calculated, the largest system size considered was $L=40$. This was reduced
to $L=28$ and $L=24$ for $\alpha =0.28$ and 0.26 respectively, due to the
increase in $2n_{\min }$. We have taken $L$ multiple of four to avoid
frustration of the next-nearest-neighbor antiferromagnetic interaction $%
J_{2}.$ 

\begin{figure}
\narrowtext
\epsfxsize=10.truecm
\vbox{\hskip 0.05truein \epsffile{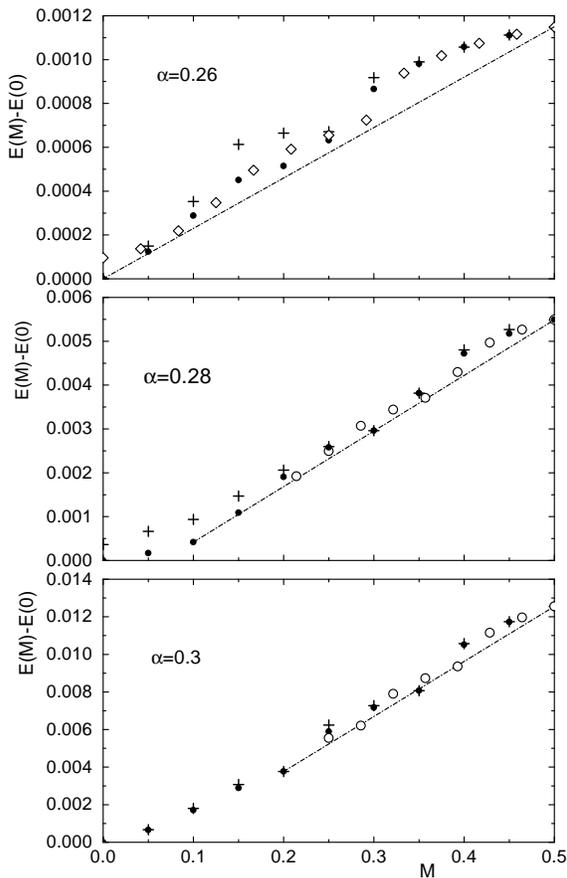}}
\medskip
\caption{Energy per site as a function of magnetization for $J_{1}=$ $%
\Delta _{2}=-\Delta _{1}=1$, and several values of $\alpha =J_{2}/J_{1}$ and
system sizes: solid circles $L=20$, empty diamonds $L=24$, and empty circles 
$L=28$. Crosses correspond to $L=20$ using periodic (instead of twisted)
boundary conditions. The dot-dashed line joins the points with $M=1/2$ and $%
M=1/2-2n_{\min }/L$ for $L=20$.}
\end{figure}

The results of $E(M)$ for two different system sizes and the three values of 
$\alpha $ are shown in Fig. 7. The oscillations with period $n_{\min }/L$
are clearly seen. We also show a comparison with the result of periodic
boundary conditions. As before, the minimization with respect to twisted
boundary conditions reduces the magnitude of these oscillations and the
finite size effects, particularly for $\alpha $ near 1/4, small $M$ and
smaller $L.$ With increasing $L$, the points tend to lie nearer to the
dot-dashed line. As mentioned before \cite{inco}, for any system size,
minimization with respect to twisted boundary conditions reproduces the
exact energy and wave vector in the one magnon sector: for $\alpha \geq 1/4$
and $M=1/2-1/L$, $q=\pm \arccos (-1/(4\alpha ))$ and 

\[
E(1/2-1/L)=(\Delta _{1}+\alpha \Delta _{2})J_{1}/4-[\Delta _{1}+\alpha
(1+\Delta _{2})+1/(8\alpha )]/L.
\]
However, for  $\alpha =0.3$, evaluation of Eq. (\ref{c2}) is not affected by
the use of periodic boundary conditions, since they minimize the energy for
three and six spin flips. Then, our results for $\alpha =0.3$ are consistent
with those of Cabra {\it et al.}, who using periodic boundary conditions
obtain that $M(B)$ is very steep near $M=1/2$, but without a jump for $%
\alpha =1/3$ (Fig. 4 of Ref. \cite{chp}). 

\begin{figure}
\narrowtext
\epsfxsize=4.0truein
\vbox{\hskip 0.05truein \epsffile{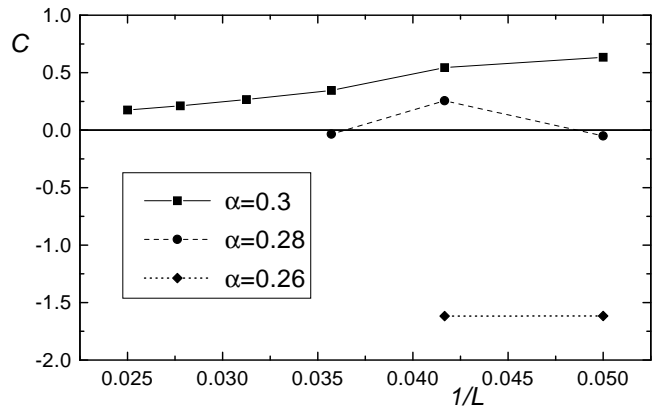}}
\medskip
\caption{Relative curvature $C=\partial ^{2}E/\partial
M^{2}|_{M=1/2}/(E(1/2)-E(0))$ calculated using Eq. (\ref{c2}) as a function
of the inverse of the system size for $J_{1}=$ $\Delta _{2}=-\Delta _{1}=1$,
and several values of $\alpha $.}
\end{figure}

In Fig. 8, we represent the dimensionless relative curvature:

\[
C=\frac{\partial ^{2}E/\partial M^{2}|_{M=1/2}}{E(1/2)-E(0)},
\]
where the numerator is evaluated using Eq. (\ref{c2}) with different values
of $L$, and in the denominator the result for $L=20$ is taken. Except for
the case of $\alpha =0.28$, in which there seems to be an oscillating
behavior of $C$ with $L$, the variation of $C$ with $L$ is rather smooth.
For  $L=24$ ($1/L\simeq 0.042$) the values of $C$ are shifted upwards with
respect to a smooth curve which fits the rest of the points. For $\alpha
=0.26$, different extrapolations of the data to the thermodynamic limit give 
$|C|<0.1$. Since this value is of the order of the uncertainty in the
extrapolations, a definite conclusion regarding the sign of $C$ cannot be
drawn. Similary, for $\alpha =0.28$, due to the oscillating behavior of the
data, no reliable extrapolation can be made. Instead, for $\alpha =0.26$,
although we have only two points available, it is remakkable that the
curvature is practically constant ($C=-1.617$) suggesting a negative value
in the thermodynamic limit. Taking into account that for $\alpha =0.28$ and
0.26, the curvature for $L=24$ seems shifted to higher values in comparison
with the rest of the curve, the sign of $C$ for all finite $L$, and the
behavior of $E(M)$ displayed in Fig. 7 for each value of $\alpha $, we
believe that $C$ changes sign for $\alpha _{c}$ slightly below $0.28$. For $%
0.25<\alpha <\alpha _{c}$, we expect a true jump in $M(B)$.

\section{Summary and discussion}

We have investigated by analytical and numerical methods, the regions of
parameters $\Delta _{1}$, $\Delta _{2}$ and $\alpha =J_{2}/J_{1}$ for which
a jump in the magnetization as a function of magnetic field $M(B)$, of the
spin-1/2 $XXZ$ chain with next-nearest-neighbor exchange (Eq. (\ref{h1})) is
expected. The numerical results are restricted to the region $|\Delta
_{1}|\leq 1$ and $0\leq \Delta _{2}\leq 1$.We were particularly interested
in parameters near the isotropic limit of ferromagnetic $J_{1}$ and
antiferromagnetic $J_{2}$ ($\alpha >0$, $\Delta _{2}=-\Delta _{1}=1$), which
are relevant to some systems containing CuO chains with edge-sharing CuO$_{4}
$ units. \cite{miz}

One of the necessary conditions for a jump in $M(B)$ is that the
ground-state energy per site as a function of magnetization $E(M)$ should
have zero or negative curvature in a finite interval. In absence of even-odd
effects or spin flips in groups as discussed in sections IV and V
respectively, the curvature at maximum $M$ ($\partial ^{2}E/\partial
M^{2}|_{M=1/2}$) can be calculated analytically in the thermodynamic limit
from the energy of the states with one and two magnons. The resulting points
of zero curvature define a surface $S_{1}$ in the parameter space $\alpha
,\Delta _{1},\Delta _{2}$, which turns out to be a frontier for the
occurrence of metamagnetic transitions. The rest of the boundary of the
region of expected metamagnetism was constructed from the condition $%
E(0)=E(1/2)$, where $E(M)$ was calculated numerically in chains of 20 sites.
We call this surface $S_{2}$.

For $-1\geq \Delta _{1}\geq 0$, the line of $S_{1}$ given by:

\[
4\alpha \Delta _{1}+1=0\text{, }1+\Delta _{2}-2\Delta _{1}^{2}=0 
\]
on which the wave vector of the two-magnon ground state in the thermodynamic
limit changes from $K=0$ to $K=\pm 2\arccos (-1/(4\alpha ))$ {\em coincides}
with a line of $S_{2}$ in which $E(M)$ is independent of $M$. This
noticeable property allows to split the regions of expected metamagnetism in
two, depending if $\alpha $ is moved to lower or higher values from this
line (see Figs. 2, 3). Inside the first region, the existence of a jump in $%
M(B)$ is confirmed calculating $E(M)$ numerically, for all possible $M$ in a
finite chain. At least for $|\Delta _{i}|\leq 1$, this region disappears if $%
\Delta _{1}=-1$ or $\Delta _{2}=1$. However, particularly for $\alpha =1/4$,
there are values of $\Delta _{1}$ and $\Delta _{2}$ near the isotropic case $%
\Delta _{2}=-\Delta _{1}=1$ for which a metamagnetic transition exists. The
intersection of $S_{1}$ with the plane $\alpha =1/4$ gives the simple
equation:

\[
2\Delta _{1}\Delta _{2}+2\Delta _{1}+3\Delta _{2}+1=0. 
\]
Displacing one of the $\Delta _{i}$ slightly to lower values from this line
(without reaching $S_{2}$) is enough to have a jump in $M(B)$ (see Fig. 1).

In the second region of expected metamagnetism (larger values of $\alpha $),
the analytical calculation of $S_{1}$ becomes invalid due to the tendency of
the system to decrease the magnetization from saturation in more than one
spin flip. Numerical calculations of the curvature $\partial ^{2}E/\partial
M^{2}$ in the isotropic case $\Delta _{2}=-\Delta _{1}=1$ near $M=1/2$ using
states with adequately chosen total spin and twisted boundary conditions
(which allow for incommensurate wave vectors \cite{inco} and are crucial
here), suggest that it remains negative for $\alpha <\alpha _{c}$ with $%
\alpha _{c}\sim 0.28$ or slightly less. A jump in $M(B)$ exists for $%
1/4<\alpha <\alpha _{c}$. In any case, even for $\alpha >\alpha _{c}$,  our
results seem enough to show that $M(B)$ should have an abrupt increase (if
not a true jump) from nearly 60\% to 100\% of the magnetization of
saturation at a critical field $B_{c}$. As an example, taking the parameters
of Ref. \cite{miz}, we estimate $B_{c}=$14 and 40 Tesla for La$_{6}$Ca$_{8}$%
Cu$_{24}$O$_{41}$ and Li$_{2}$CuO$_{2}$ respectively.

\section*{Acknowledgments}

I am grateful to F.H.L. E\ss ler, C.D. Batista, D. Cabra, A. Honecker and
Ana L\'{o}pez and for important discussions, and to Karen Hallberg for a
critical reading of the manuscript. I acknowledge computer time at the
Max-Planck Institute f\"{u}r Physik Komplexer Systeme. I am partially
supported by CONICET. This work was sponsored by PICT 03-00121-02153 of
ANPCyT and PIP 4952/96 of CONICET.

\end{document}